\documentstyle{article}

\newcommand{\four}{\mathop{{{}^{{}^{(4)}}}\!\!}}
\newcommand{\three}{\mathop{{{}^{{}^{(3)}}}\!\!}}
\newcommand{\diag}{\mathop {\rm diag}\nolimits}
\newcommand{\dfrac}[2]{\frac{\displaystyle{#1}}{\displaystyle{#2}}}

\begin{document}

\title{Dirac Decomposition of Wheeler-DeWitt Equation\\
on the Bianchi Class A Models}

\author{Hidetomo {\sc Yamazaki} 
\thanks{E-mail address: yamazaki@cc.kyoto-su.ac.jp}\\[3mm]
{\it Department of Physics, Kyoto Sangyo University, Kyoto 
603-8555, Japan}}

\date{}

\maketitle

\begin{abstract}
  The Wheeler-DeWitt equation for the Bianchi Class A
  cosmological models is expressed generally in terms of the
  second-order differential equation like the Klein-Gordon
  equation. To obtain the positive-definite probability
  density, a new method extending the Dirac-Square-Root
  formalism, which factorizes the Wheeler-DeWitt equation
  into the first-order differential equation using the Pauli
  matrices, is investigated.  The solutions to the Dirac
  type equation thus obtained are expressed in terms of
  two-component spinor form. The probability density defined
  by the solution is positive-definite and there is a
  conserved current. The newly found spin-like degree of
  freedom causes the universe to go through an early quantum
  stage of evolution with agitated anisotropy-oscillation
  like Zitterbewegung.
\end{abstract}

\section{Introduction}

Arnowitt-Deser-Misner \cite{rf:ADM}(ADM) reformulated the general
relativity using the canonical formalism, which makes a 3+1 split of
the space-time metric, and introduced some constraints through the
variational principle.  According to the ADM approach, the Einstein
equation is derived as the equation of dynamical evolution.  The
canonical formalism by the ADM approach is a system of dynamics
including two constraints: one is the momentum constraint and the
other is the Hamiltonian constraint. The latter governs the time
evolution of the system.  According to Dirac \cite{rf:Dirac} the
procedure of canonical quantization replaces the Hamiltonian
constraint with the supplementary condition on the wave function which
represents the quantum state of space-time.  The Hamiltonian
constraint thus becomes the Wheeler-DeWitt equation, which is
fundamental to the quantum gravity and determines the quantum state of
space-time.  The Wheeler-DeWitt equations for specific cosmological
models like the Friedmann-Robertson-Walker model  \cite{rf:DeWitt} or
the Bianchi models \cite{rf:Misner,rf:Ryan72,rf:Ryan-Shepley} have been
investigated.

The Wheeler-DeWitt equation for the Bianchi models is in
general reduced to the second-order differential equation if
the operator-ordering is neglected. The resultant
Wheeler-DeWitt equation is similar to the Klein-Gordon type,
which has a problem that the probability density is not
positive-definite.  The Wheeler-DeWitt equation has also
often this kind of problem that the probability density
becomes negative. The negative values of probability density
have been shown by Furusawa \cite{rf:Furusawa} through
numerical calculation, who treated the Wheeler-DeWitt
equation for the quantum mixmaster model using wave-packet solution.

To avoid the problem, the approach called Dirac-Square-Root
formalism has been
studied  \cite{rf:Ryan72,rf:Ryan-Shepley,rf:Mallett,rf:Kim-Oh}.
Since the Hamiltonian for the Bianchi type models is
expressed in terms of the quadratic, the first-order
differential equation can be derived by applying a procedure
similar to the Dirac method.  Mallett \cite{rf:Mallett}
applied it to the Friedmann model coupled to the charged
scalar field. Kim and Oh \cite{rf:Kim-Oh} mainly discussed
the Bianchi type-IX model.
In their papers, the first-order equation included an
unknown function dependent on the time-parameter is
postulated and the wave function is required to be a
solution of two-component column vector.  Iterating the
first-order equation thus postulated, the differential
equation with respect to the unknown function is yielded so
as to be consistent with the Wheeler-DeWitt equation for
each component.  The differential equation with respect to
the time parameter is thus obtained. 
Indeed, the differential equation is reduced to Riccati or
Bernoulli equation in case of the Friedmann model with charged
scalar field or the Bianchi type-IX model, respectively.
However, one sees that it is difficult to obtain the exact
solutions in both cases.

The purpose of this paper is to propose a new method, which
factorizes the Wheeler-DeWitt equation into the first-order
differential equation, using the Pauli matrices. The
resultant differential equation thus obtained is similar
to the Dirac equation. As a necessary condition for the
Dirac factorization, it is postulated that the resultant
Hamiltonian is ``self-adjoint''. We apply this method to the
the Wheeler-DeWitt equation of the Bianchi Class A vacuum
models except for type-IX model. The solution of the Dirac
type equation is written by two-component column
vector. Using the Dirac type equation and the complex
conjugate, we can define the probability density. Then, we
get the equation of continuity and the conserved
current. Here, we restrict ourselves to determine the form
of Hamiltonian so as to make the probability density
positive-definite. It is to be noted that our formalism
cannot be applied to the Bianchi type-IX model, because the
structure of anisotropic potential for the Bianchi type-IX
model is different from that of the other Bianchi Class A
models.

The organization of this paper is as follows.  In \S 2, we
review the ADM canonical formulation for the vacuum Bianchi
Class A models and introduce the Wheeler-DeWitt equation.
In \S 3, we propose the new method in which the
Wheeler-DeWitt equation is factorized into the first-order
differential equation in terms of the Pauli matrices and
apply the method to the Bianchi type-I, II, VI$_0$, VII$_0$
and VIII models.  For each of them, the Dirac type equation
is derived and solved except for the Bianchi type-VIII
model. Then, we assume a trail function which is expressed
in terms of two-component spinor. We show that the
probability density becomes positive-definite and there is a
conserved current. As the time
parameter approaches the limit of the singularity of the
universe, the Wheeler-DeWitt equation for the Bianchi type-II,
VI$_0$, VII$_0$ and VIII models are reduced to that of
the Bianchi type-I model, so that the Dirac type equations 
must possess the solution for the Bianchi type-I model.
In \S 4, we discuss the behavior for the Dirac type equation
near the singularity in detail. 
It is shown that from the Heisenberg equation the quantum behavior
of the anisotropic parameters $\beta_\pm$ make the quivering motion
and the Hamiltonian system has a `spin-like' degree of freedom. The
`spin-like' degree of freedom causes the universe to go
necessarily through an early quantum stage of evolution with
agitated anisotropy-oscillation.  In \S 5, summary and
discussion are made on the results of our formalism and
comparisons are made with the Dirac-Square-Root formalism.
Henceforward, we take that $i$, $j$ and $k$ runs from 1 to 3
and the unit $c=\hbar=16\pi G=1$.

\section{Canonical Quantization formulation for the Bianchi
  Class A models}

The Einstein-Hilbert action is given
by \cite{rf:MTW}
\begin{equation}
  I = \int \sqrt{-\four g} \four R \ d^4x,
  \label{eq:action}
\end{equation}
where $\four g$ and $\four R$ are the determinant of the
space-time metric and the curvature, respectively; the
indices $(4)$ implies 4-geometry quantity.  The metric for
the Bianchi models is generally given by
\begin{equation}
  ds^2 = -N^2 dt^2 + e^{2\alpha} e^{2\beta_{ij}}
  \chi^i \chi^j,\label{eq:metric}
\end{equation}
where $N$, $\alpha$ and $\beta_{ij}$ are functions of $t$
only: $N$ is the lapse function; $e^{2\alpha} $ plays the
role of a scale factor, so that the universe has the initial
singularity as $\alpha\rightarrow -\infty$; $\beta_{ij}$ are
an isotropic parameters of the universe, which gives a matrix
form of traceless and $\chi^i$ are 1-forms. According to
Misner \cite{rf:Misner}, anisotropic parameters $\beta_{ij}$
are expressed in terms of two parameters $\beta_+$ and
$\beta_-$
\begin{equation}
  \beta_{ij} = \diag (\beta_+ + \sqrt{3}\beta_-, \beta_+ -
  \sqrt{3}\beta_-, -2\beta_+).
  \label{eq:param-bb}
\end{equation}
The 1-forms obey
\begin{equation}
  d\chi^i = \frac{1}{2} C^i_{jk} \chi^j \wedge \chi^k,
  \label{eq:M-C}
\end{equation}
where $C^i_{jk}$ are the structure constants. 

The Bianchi models are characterized by the structure
constants.  The Bianchi models are classified into two
groups from the structure constants, so-called Class A and
B \cite{rf:Ellis-Mac}. It should be noted that, as was first
indicated by Hawking \cite{rf:Hawking}, it cannot be reduced
to the correct field equations by the ADM canonical
formulation in the Bianchi Class B models except for the
Bianchi type-V model \cite{rf:Ryan74,rf:Sneddon}.  The reason
is that the spatial divergence term does not vanish. Hence,
we must discuss only the Bianchi Class A models because we
use the canonical formulation.  The characteristic of the
Bianchi Class A models is that the structure constants obey
the condition
\begin{equation}
  {C^i}_{ji} = 0. \label{eq:class-A}
\end{equation}
We must therefore select the 1-forms so as to satisfy
Eq. (\ref{eq:class-A}).

After the procedure for the ADM decomposition, we use the
coordinates $\alpha$, $\beta_+$, $\beta_-$ and the
corresponding canonical momenta as $p_\alpha, p_+,
p_-$. From the variation of $N$, one obtains the Hamiltonian
constraint.  The 3-geometry curvature $\three R$ is obtained
from the structure constants.  According to
Misner \cite{rf:Misner}, the relation of the curvature
$\three R$ to the anisotropic potential $V(\beta_+,
\beta_-)$ is introduced by
\begin{equation}
  \three R = \frac{3}{2} e^{2\alpha} \left( 1-V(\beta_+,
    \beta_-) \right)
  \label{eq:3R-V}.
\end{equation}
As a result, the Hamiltonian constraint becomes
\begin{equation}
  H = -p_\alpha^2 + p_+^2 + p_-^2 + e^{4\alpha}(V-1).
\end{equation}

On the canonical quantization procedure, we treat the
operators $p_\alpha$, $p_+$, $p_-$ in the Schr\"odinger
representation:
\begin{equation}
  \hat{p}_\alpha = -i \frac{\partial}{\partial\alpha}, \quad
  \hat{p}_+ = - i \frac{\partial}{\partial\beta_+}, \quad
  \hat{p}_- = - i \frac{\partial}{\partial\beta_-}.
  \label{eq:diff-op}
\end{equation}
Because the Hamiltonian constraint is the first-class
constraint, the quantized Hamiltonian constraint must be
satisfied by the supplementary condition
\begin{equation}
  \hat{H} \Psi = 0 \label{eq:WW},
\end{equation}
\begin{equation}
  \hat{H} \equiv
  \frac{\partial^2}{\partial\alpha^2}
  - \frac{\partial^2}{\partial\beta_+^2}
  - \frac{\partial^2}{\partial\beta_-^2}
  + e^{4\alpha}(V-1)
  \label{eq:hamiltonian},
\end{equation}
where $\Psi = \Psi(\alpha, \beta_+,\beta_-)$ is a state
vector of the system, called by the wave function of the
universe.  This is the Wheeler-DeWitt equation for the
Bianchi models.

\section{Dirac factorization for the Wheeler-DeWitt equation}

According to the Furusawa's discussion \cite{rf:Furusawa},
the probability density of Klein-Gordon type
\begin{eqnarray*}
  \rho(\alpha, \beta_+, \beta_-) 
  =
  \frac{i}{2}
  \left( \Psi^* \partial_\alpha \Psi - \Psi
    \partial_\alpha \Psi^*
  \right)
\end{eqnarray*}
becomes negative for some values $\alpha$. The negative
values of the probability density are inconsistent with the
physical theories and it is the failure of the structure in
the Wheeler-DeWitt equation. To get rid of the problem, we
adopt the similar procedure that Dirac factorized the
Klein-Gordon equation in order to obtain the
positive-definite probability density. That is, we try to
factorize the Wheeler-DeWitt equation into the first-order
differential equation. We denote the first-order equation as
Dirac type equation.  As a necessary condition of Dirac
factorization, it is postulated that the resultant
Hamiltonian is ``self-adjoint'', because one tries to think
various ways of the factorization into the first-order
equation and there is not any rules for the factorization.
We suppose that the Wheeler-DeWitt equation for the Bianchi
models is replaced by the equation of Dirac type
\begin{equation}
  {\cal H} \Psi
  \equiv
  \left(
  \hat{p}_\alpha
  + \sigma_1 \hat{p}_+
  + \sigma_2 \hat{p}_-
  + \sigma_3 e^{2\alpha} v
  \right) \Psi
  = 0,
  \label{eq:Dirac}
\end{equation}
where $v \equiv \sqrt{V - 1}$ and the Pauli matrices
$\sigma_i$ are expressed as
\begin{eqnarray*}
  \sigma_1 =
  \left(
    \begin{array}{cc}
      0 & 1 \\
      1 & 0
    \end{array}
  \right),   \quad
  \sigma_2 =
  \left(
    \begin{array}{cc}
      0 & -i \\
      i & 0
    \end{array}
  \right) , \quad
  \sigma_3 =
  \left(
    \begin{array}{cc}
      1 & 0 \\
      0 & -1
    \end{array}
  \right).
\end{eqnarray*}
We here rewrite the Dirac type equation (\ref{eq:Dirac}) in
the Schr\"odinger form
\begin{equation}
  i \frac{\partial}{\partial\alpha} \Psi
  =
  \left\lbrack
    \sigma_1 \hat{p}_+
    + \sigma_2 \hat{p}_-
    + \sigma_3 e^{2\alpha} v
  \right\rbrack
  \Psi.
\end{equation}

The self-adjointness of the Hamiltonian is easily proved:
\begin{equation}
  {\cal H}^\dagger
  =
  \hat{p}_\alpha
  + \sigma_1 \hat{p}_+
  + \sigma_2 \hat{p}_-
  + \sigma_3 e^{2\alpha} v
  = {\cal H},
\end{equation}
where the momentum operators are self-adjoint and commute
with the Pauli matrices.

With the help of relations (\ref{eq:diff-op}), the Dirac
type equation (\ref{eq:Dirac}) is rewritten as
\begin{equation}
  -i \frac{\partial\Psi}{\partial\alpha}
  -i \sigma_1 \frac{\partial\Psi}{\partial\beta_+}
  -i \sigma_2 \frac{\partial\Psi}{\partial\beta_-}
  + \sigma_3 e^{2\alpha} v \Psi = 0. \label{eq:Dirac-type}
\end{equation}
Taking the complex conjugate and multiplying $\Psi^\dagger$
to Eq.  (\ref{eq:Dirac-type}) from the left-hand side, we
have
\begin{equation}
   i \Psi^\dagger \frac{\partial\Psi}{\partial\alpha}
  + i \Psi^\dagger \sigma_1 \frac{\partial\Psi}{\partial\beta_+}
  + i \Psi^\dagger \sigma_2 \frac{\partial\Psi}{\partial\beta_-}
  +  \Psi^\dagger  \sigma_3 e^{2\alpha} v \Psi = 0.
  \label{eq:p1-p}
\end{equation}
Subtracting Eq. (\ref{eq:p1-p}) from
Eq. (\ref{eq:Dirac-type}) multiplied $\Psi^\dagger$ from the
left-hand side, we obtain
\begin{equation}
  \frac{\partial}{\partial\alpha} \Psi^\dagger\Psi
  + \frac{\partial}{\partial\beta_+}
  \left( \Psi^\dagger \sigma_1 \Psi \right)
  + \frac{\partial}{\partial\beta_-}
  \left( \Psi^\dagger \sigma_2 \Psi \right)
  = 0.
\end{equation}
Setting the probability density $\rho$ and the current
vectors $j_+$ and $j_-$ as
\begin{equation}
  \rho = \Psi^\dagger\Psi, \quad
  j_+ = \Psi^\dagger \sigma_1 \Psi, \quad
  j_- = \Psi^\dagger \sigma_2 \Psi,
  \label{eq:current}
\end{equation}
we get the equation of continuity
\begin{equation}
  \frac{\partial\rho}{\partial\alpha}
  + \frac{\partial j_+}{\partial\beta_+}
  + \frac{\partial j_-}{\partial\beta_-}
  = 0. \label{eq:continuity}
\end{equation}
Then, one sees that the probability density is
positive-definite.  We apply our formalism to solve the
Bianchi type-I, II, VI$_0$, VII$_0$ and VIII models below.

\subsection{The Bianchi type-I model}

The metric of the Bianchi type-I model is given in
Eq. (\ref{eq:metric}), in which the 1-forms are given
by \cite{rf:MacCallum}
\begin{equation}
  \chi^1 = dx^1, \quad
  \chi^2 = dx^2, \quad
  \chi^3 = dx^3 \label{eq:I-basis}.
\end{equation}
The structure constants in the model obviously supply
\begin{equation}
  {C^i}_{jk} = 0.
\end{equation}
It leads to the zero curvature of 3-dimensional
hypersurfaces $\three R_{\rm I} = 0$, which implies the
spatial flat structure. From this, the anisotropic potential
becomes $V_{\rm I} =1$. Then, the Wheeler-DeWitt equation is
reduced to
\begin{equation}
  \label{eq:W-W1}
  \left(
    - \hat{p}_\alpha^2
    + \hat{p}_+^2
    + \hat{p}_-^2
    \right) \Psi = 0.
\end{equation}

We postulate here that the Dirac type equation for the
Bianchi type-I model holds:
\begin{equation}
  {\cal H}_{\rm I} \Psi_{\rm I} (\alpha, \beta_+, \beta_-)
  =
  \left(
    \hat{p}_\alpha 
    + \sigma_1 \hat{p}_+ 
    + \sigma_2 \hat{p}_-
  \right) \Psi
  = 0 \label{eq:dirac-type1}.
\end{equation}
The solutions to this equation have been already
found \cite{rf:Ryan72,rf:Ryan-Shepley}. They are plane wave
solutions given by two-component spinor form
\begin{equation}
  \Psi_{\rm I}(\alpha, \beta_+, \beta_-)
   =
   e^{i ( k_+ \beta_+ + k_- \beta_- - \omega \alpha)}
  \left(
    \begin{array}{c}
      \psi_1 \\ \psi_2
    \end{array}
  \right)
  \label{eq:plane-wave},
\end{equation}
where $\omega$ and $k_\pm$ are arbitrary real constants.
Substituting the solution (\ref{eq:plane-wave}) into the
Dirac type equation (\ref{eq:dirac-type1}), we get the
components
\begin{equation}
  \psi_1 = {k_+ - ik_- \over \omega} \psi_2 
  \quad {\rm and} \quad
  \psi_2 = {k_+ + ik_- \over \omega} \psi_1.
\end{equation}
It implies obviously $\vert\psi_1\vert^2 = \vert\psi_2\vert^2$
and the relation among $\omega$ and $k_\pm$ is
\begin{equation}
  \omega = \pm \sqrt{k_+^2 + k_-^2} .
\end{equation}
Thus, the parameter $\omega$ corresponding to the frequency
of the plane wave depends on the parameters $k_\pm$
corresponding to the wave number, which vary continuously.

The probability density is
\begin{equation}
  \rho_{\rm I}(\alpha, \beta_+, \beta_-)
  = \Psi_{\rm I}^\dagger \Psi_{\rm I} 
  = \vert \psi_1 \vert^2 + \vert \psi_2 \vert^2 
  = 2 \vert \psi_1 \vert^2,
\end{equation}
which is positive-definite. The current vectors $j_\pm$ are
respectively
\begin{equation}
  j_+ = \psi_2^*\psi_1 + \psi_1^*\psi_2
  \label{eq:type-I-current+}
\end{equation}
and
\begin{equation}
  j_- = i \psi_2^*\psi_1 -i\psi_1^*\psi_2,
  \label{eq:type-I-current-}
\end{equation}
which take some constants.  Therefore, the equation of
continuity (\ref{eq:continuity}) holds and there is a
conserved current. From the parameters (\ref{eq:param-bb})
and the result of Eq. (\ref{eq:type-I-current+}) and
(\ref{eq:type-I-current-}), we interpret that the evolution
of the quantized universe is that whether one of the three
spatial axes is expanding and two axes are contracting or
one axis is contracting and two axes are expanding.

It should be noted that the quantity
\begin{equation}
  \int \rho_{\rm I}(\alpha, \beta_+, \beta_-) d\beta_+
  d\beta_-
  = 2 \vert \psi_1 \vert^2 \int d\beta_+d\beta_-
  \label{eq:probability}
\end{equation}
is not convergent and the wave function
(\ref{eq:plane-wave}) is not the square integrable function.
Therefore, it is impossible to normalize the wave function
(\ref{eq:plane-wave}) without an appropriate boundary
condition and we do not discuss the normalization constant
in this paper. Here, the quantity (\ref{eq:probability})
does not directly determine the probability, so that the
probability must be regarded as a quantity proportional to
this quantity (\ref{eq:probability}).  Moreover, since
$\partial\rho/\partial\alpha = 0$, the probability density
is independent of the time-parameter and quantity
(\ref{eq:probability}) is conserved.

As mentioned above, it should be noted here that since the
Wheeler-DeWitt equation and the Dirac type equation for the
Bianchi type-II, VI$_0$, VII$_0$ and VIII models are reduced
to that of the Bianchi type-I model at the singularity
($\alpha\rightarrow -\infty$), all models have the same
property that the behavior near the singularity becomes that
of the plane wave solution. We shall discuss the quantized
Hamiltonian system of the Bianchi type-I model in \S 4.

\subsection{The Bianchi type-II model}

The metric of the Bianchi type-II model is expressed as
Eq. (\ref{eq:metric}) and the 1-forms are given
by \cite{rf:MacCallum}
\begin{equation}
  \chi^1 = dx^1 - x^3 dx^2, \quad
  \chi^2 = dx^2, \quad
  \chi^3 = dx^3 \label{eq:II-basis}.
\end{equation}
Substituting the 1-forms (\ref{eq:II-basis}) into
Eq. (\ref{eq:M-C}), we get the structure constants
\begin{equation}
  C^1_{23} = - C^1_{32} = 1.
\end{equation}
Other components are zero. The curvature is
\begin{equation}
  \three R_{\rm II}
  = 
  - \frac{1}{2} e^{2\alpha + 4\beta_+ + 4\sqrt{3}\beta_-}
  \label{eq:3curvature},
\end{equation}
and the anisotropic potential $V_{\rm II}(\beta_+, \beta_-)$
is obtained from the relation (\ref{eq:3R-V}) as
\begin{equation}
  V_{\rm II} - 1 = \frac{1}{3} e^{4\beta_+ +
    4\sqrt{3}\beta_-}.
\end{equation}
Then, the Wheeler-DeWitt equation is expressed as
\begin{equation}
  \label{eq:W-W2}
  \left(
    - \hat{p}_\alpha^2
    + \hat{p}_+^2
    + \hat{p}_-^2
    + \frac{1}{3}e^{4\alpha + 4\beta_+ + 4\sqrt{3}\beta_-}
    \right) \Psi = 0.
\end{equation}
We postulate here that the factorized Hamiltonian is
expressed as
\begin{equation}
  {\cal H}_{\rm II}
  =
  \hat{p}_\alpha 
  + \sigma_1 \hat{p}_+ 
  + \sigma_2 \hat{p}_- 
  + \frac{\sigma_3}{\sqrt{3}} e^{2\alpha + 2\beta_+ +
    2\sqrt{3} \beta_-}
  \label{eq:H-II}
\end{equation}
and that the Dirac type equation holds:
\begin{equation}
  {\cal H}_{\rm II} \Psi_{\rm II} = 0.
  \label{eq:Dirac-2a}
\end{equation}
With the help of the relation (\ref{eq:diff-op}), it becomes
\begin{equation}
  i \frac{\partial \Psi_{\rm II}}{\partial\alpha}
  =
  \left(
    \begin{array}{cc}
      \dfrac{1}{\sqrt{3}} 
      e^{2\alpha + 2\beta_+ + 2\sqrt{3} \beta_-} &
      -i \dfrac{\partial}{\partial\beta_+}
      - \dfrac{\partial}{\partial\beta_-} \\[3mm]
      -i \dfrac{\partial}{\partial\beta_+}
      + \dfrac{\partial}{\partial\beta_-} &
      - \dfrac{1}{\sqrt{3}} 
      e^{2\alpha + 2\beta_+ + 2\sqrt{3} \beta_-}
    \end{array}
  \right)
  \Psi_{\rm II} .
  \label{eq:schroedinger}
\end{equation}

Now we look for a solution to this equation. We take a trial
function which gives the two-component spinor form and
notices the form of the Dirac type equation as
\begin{equation}
  \Psi_{\rm II}
  = \exp \left\lbrack
    - \frac{1}{6} e^{2\alpha + 2\beta_+ + 2\sqrt{3}\beta_-}
    \right\rbrack
    \left(
      \begin{array}{c}
        \psi_1 \\ \psi_2
      \end{array}
    \right), \label{eq:II-trial}
\end{equation}
where the spinor components $\psi_1$ and $\psi_2$ are
arbitrary constants. Substituting the trail function
(\ref{eq:II-trial}) into Eq. (\ref{eq:schroedinger}), we
get the relation of the spinor components
\begin{equation}
  \psi_2 = - \psi_1\label{eq:spinor-H1}.
\end{equation}

Here we use the condition that the Dirac type equation
(\ref{eq:Dirac-2a}) for the Bianchi type-II model is reduced
to that for the Bianchi type-I model as the limit of the
singularity ($\alpha \rightarrow -\infty$).  The trail
function (\ref{eq:II-trial}) would include such an
asymptotic behavior.  The trail function involving the plane
wave part is expressed as
\begin{equation}
  \Psi_{\rm II} 
  =
  e^{i(k_+ \beta_+ + k_-\beta_- - \omega\alpha)}
  \exp \left\lbrack
    - \frac{1}{6} e^{2\alpha + 2\beta_+ + 2\sqrt{3}\beta_-}
  \right\rbrack
  \left( \begin{array}{c}
      \psi_1 \\ - \psi_1
    \end{array}
  \right)  \label{eq:wavefunction}.
\end{equation}
Although $\omega$, $k_\pm$ are arbitrary real constants, the
coefficients $\omega$ and $k_\pm$ are not free at all.
Substituting the solution (\ref{eq:wavefunction}) into
Eq. (\ref{eq:schroedinger}), we have
\begin{equation}
  \left( \begin{array}{cc}
      - \omega & k_+ -ik_- \\
      k_+ + ik_- & - \omega
  \end{array} 
  \right)
  \left( \begin{array}{c}
      \psi_1 \\ - \psi_1
  \end{array} \right)
  = 0.
\end{equation}
Therefore, the coefficients must satisfy
\begin{equation}
  k_+ = - \omega , \quad  k_- = 0 \label{eq:sindou}.
\end{equation}
Consequently, the solution is finally expressed in terms of
the product of the amplitude and oscillation parts
restricted by Eq. (\ref{eq:sindou}):
\begin{equation}
  \Psi_{\rm II} 
  =
  e^{- i \omega (\beta_+ + \alpha)}
  \exp \left\lbrack
    - \frac{1}{6} e^{2\alpha + 2\beta_+ + 2\sqrt{3}\beta_-}
  \right\rbrack
  \left( \begin{array}{c}
      \psi_1 \\ - \psi_1
    \end{array}
  \right)  \label{eq:solution-2}.
\end{equation}

The probability density is expressed, by using the wave
function (\ref{eq:solution-2}), as
\begin{equation}
  \rho_{\rm II}(\alpha,\,\beta_+,\,\beta_-) 
  = \Psi^\dagger_{\rm II} \Psi_{\rm II}
  = 2 \vert \psi_1 \vert^2
  \exp
  \left\lbrack
    - \frac{1}{3} e^{2\alpha + 2\beta_+ + 2\sqrt{3}\beta_-}
  \right\rbrack
  \label{eq:2-probability}.
\end{equation}
This is always positive-definite.  One sees that this
probability density goes to zero as $\beta_+ +
\sqrt{3}\beta_- \rightarrow \infty$, while it approaches a
certain constant as $\beta_+ + \sqrt{3}\beta_- \rightarrow -
\infty$. The current vectors $j_\pm$ given by
Eq. (\ref{eq:current}) are
\begin{equation}
  \label{eq:j+2}
  j_+ = -2 \vert \psi_1 \vert^2 
  \exp
  \left\lbrack
    - \frac{1}{3} e^{2\alpha + 2\beta_+ + 2\sqrt{3}\beta_-}
  \right\rbrack
\end{equation}
and
\begin{equation}
  \label{eq:j-2}
  j_- = 0.
\end{equation}
Hence, the equation of continuity (\ref{eq:continuity})
holds and there is the conserved current. Moreover, it is
shown that the wave part propagates to the direction of
$\beta_+$ axis only and it agrees with the result of
Eq. (\ref{eq:sindou}).

\subsection{The Bianchi type-VI$_0$ model}

The metric of the Bianchi type-VI$_0$ model is expressed as
Eq.  (\ref{eq:metric}) and the 1-forms are given
by \cite{rf:MacCallum}
\begin{equation}
  \chi^1 = \cosh x^3 dx^1 - \sinh x^3 dx^2 ,\quad
  \chi^2 = - \sinh x^3 dx^1 + \cosh x^3 dx^2 ,\quad
  \chi^3 = dx^3. \label{eq:VI-basis}
\end{equation}
The structure constants are
\begin{equation}
  C^1_{23} = - C^1_{32} = 1 ,\quad
  C^2_{31} = - C^2_{13} = -1
  \label{eq:str-c-6}
\end{equation}
and the rest are zero.  The anisotropic potential becomes
\begin{equation}
  V_{\rm VI_0} - 1 = \frac{4}{3} e^{4\beta_+} 
  \left( \cosh(4\sqrt{3}\beta_-) +1
  \right).
\end{equation}
Then, the Wheeler-DeWitt equation for the Bianchi
type-VI$_0$ model is expressed as
\begin{equation}
  \label{eq:W-W6}
  \left\lbrack
    - \hat{p}_\alpha^2
    + \hat{p}_+^2
    + \hat{p}_-^2
    + \frac{4}{3} e^{4\beta_+} 
    \left( \cosh(4\sqrt{3}\beta_-) +1
    \right)
  \right\rbrack \Psi = 0.
\end{equation}

We postulate that the Dirac type equation for the Bianchi
type-VI$_0$ model is expressed as
\begin{equation}
  {\cal H}_{\rm VI_0} \Psi_{\rm VI_0} 
  \equiv 
  \left\lbrack
    \hat{p}_\alpha 
    + \sigma_1 \hat{p}_+ 
    + \sigma_2 \hat{p}_- 
    + \frac{2}{\sqrt{3}} \sigma_3 e^{2\alpha + 2\beta_+}
    \left( e^{2\sqrt{3}\beta_-} + e^{-2\sqrt{3}\beta_-}
    \right)
  \right\rbrack \Psi_{\rm VI_0} = 0.
  \label{eq:Dirac-6}
\end{equation}
As in the case of the Bianchi type-II model, we assume
 a trial function of the form
\begin{equation}
  \label{eq:trail-solution-6}
  \Psi_{\rm VI_0} = \exp\left\lbrack
    - \frac{1}{3} e^{2\alpha + 2\beta_+}
    \left(
      e^{2\sqrt{3}\beta_-} - e^{-2\sqrt{3}\beta_-}
    \right)
  \right\rbrack
  \left(
    \begin{array}{c}
      \psi_1 \\ \psi_2
    \end{array}
  \right),
\end{equation}
where the components $\psi_1$ and $\psi_2$ are arbitrary
constants. Substituting the trial function
(\ref{eq:trail-solution-6}) into the Dirac type equation
(\ref{eq:Dirac-6}), we get
\begin{equation}
  \psi_2 = - \psi_1 
  \label{eq:spinor-6}.
\end{equation}
From the condition that the wave function
(\ref{eq:trail-solution-6}) should be reduced to the plane
wave solution at the limit of the singularity, we obtain
\begin{equation}
  k_+ = - \omega , \quad  k_- = 0.
\end{equation}
The wave function involving the plane wave part is finally
expressed as
\begin{equation}
  \Psi_{\rm VI_0}
  = 
  e^{- i \omega (\beta_+ + \alpha)}
  \exp\left\lbrack
    - \frac{1}{3}
    e^{2\alpha + 2\beta_+}
    \left(
      e^{2\sqrt{3}\beta_-} - e^{-2\sqrt{3}\beta_-}
    \right)
  \right\rbrack
  \left(
    \begin{array}{c}
      \psi_1 \\ -\psi_1
    \end{array}
  \right). \label{eq:solution-6}
\end{equation}

The probability density becomes
\begin{equation}
  \rho_{\rm VI_0}(\alpha, \beta_+, \beta_-)
  = \Psi_{\rm VI_0}^\dagger \Psi_{\rm VI_0}
  = 2 \vert \psi_1 \vert^2
  \exp
  \left\lbrack
    - \frac{2}{3} e^{2\alpha + 2\beta_+}
     \left(
       e^{2\sqrt{3}\beta_-} - e^{-2\sqrt{3}\beta_-}
    \right)
  \right\rbrack
  \label{eq:6-probability},
\end{equation}
which is positive-definite.  The current vectors are
\begin{equation}
  j_+ = 
  -2 \vert \psi_1 \vert^2
  \exp
  \left\lbrack
    - \frac{2}{3} e^{2\alpha + 2\beta_+}
    \left(
      e^{2\sqrt{3}\beta_-} - e^{-2\sqrt{3}\beta_-}
    \right)
  \right\rbrack,
  \quad
  j_- = 0.
\end{equation}
One sees that the equation of continuity holds and there is
a conserved current.  The plane wave part of the wave
function (\ref{eq:solution-6}) propagates to the direction
of $\beta_+$ axis only. It is to be noted that since the wave
function (\ref{eq:solution-6}) and the probability density
(\ref{eq:6-probability}) are both divergent as $\beta_-
\rightarrow - \infty$, this solution is unphysical in such a
region.

\subsection{The Bianchi type-VII$_0$ model}

The metric of the Bianchi type-VII$_0$ model is expressed as
Eq.  (\ref{eq:metric}) and the 1-forms are given
by \cite{rf:MacCallum}
\begin{equation}
  \chi^1 = \cos x^1 dx^2 + \sin x^1 dx^3 ,\quad
  \chi^2 =  - \sin x^1 dx^2 + \cos x^1 dx^3 , \quad
  \chi^3 = dx^1.
  \label{eq:VII-basis}
\end{equation}
The structure constants are
\begin{equation}
  C^1_{23} = - C^1_{32} = -1 ,\quad
  C^2_{31} = - C^2_{13} = -1
  \label{eq:str-c-7}
\end{equation}
and the rest are zero.  The anisotropic potential is
\begin{equation}
  V_{\rm VII_0} - 1 = \frac{4}{3} e^{4\beta_+}
  \left(
    \cosh(4\sqrt{3}\beta_-) -1
  \right) .
\end{equation}
Then, the Wheeler-DeWitt equation for the Bianchi
type-VII$_0$ model is expressed as
\begin{equation}
  \label{eq:W-W7}
  \left\lbrack
    \hat{p}_\alpha^2
    + \hat{p}_+^2
    + \hat{p}_-^2
    + \frac{4}{3} e^{4\beta_+} 
    \left( \cosh(4\sqrt{3}\beta_-) -1
    \right)
  \right\rbrack \Psi = 0.
\end{equation}
Here, we postulate that the Dirac type equation with a
factorized Hamiltonian is expressed as
\begin{equation}
  {\cal H}_{\rm VI_0} \Psi_{\rm VII_0} 
  \equiv
  \left\lbrack
    - \hat{p}_\alpha 
    + \sigma_1 \hat{p}_+ 
    + \sigma_2 \hat{p}_- 
    + \frac{2}{\sqrt{3}} \sigma_3 e^{2\alpha + 2\beta_+}
    \left( e^{2\sqrt{3}\beta_-} - e^{-2\sqrt{3}\beta_-}
    \right)
  \right\rbrack \Psi_{\rm VII_0} = 0.
  \label{eq:Dirac-7}
\end{equation}
As in the case of previous models, we can find out the
wave function
\begin{equation}
  \Psi_{\rm VII_0} = 
  e^{- i \omega (\beta_+ + \alpha)}
  \exp\left\lbrack
    - \frac{1}{3}
    e^{2\alpha + 2\beta_+}
    \left(
      e^{2\sqrt{3}\beta_-} + e^{-2\sqrt{3}\beta_-}
    \right)
  \right\rbrack
  \left(
    \begin{array}{c}
      \psi_1 \\ -\psi_1
    \end{array}
  \right) \label{eq:solution-7}.
\end{equation}

The probability density becomes
\begin{equation}
  \rho_{\rm VII_0}(\alpha, \beta_+, \beta_-)
  = \Psi_{\rm VII_0}^\dagger \Psi_{\rm VII_0}
  = 2 \vert \psi_1 \vert^2
  \exp
  \left\lbrack
    - \frac{2}{3} e^{2\alpha + 2\beta_+}
     \left(
       e^{2\sqrt{3}\beta_-} + e^{-2\sqrt{3}\beta_-}
    \right)
  \right\rbrack
  \label{eq:7-probability},
\end{equation}
which is positive-definite. The current vectors have
\begin{equation}
  j_+ = 2 \vert \psi_1 \vert^2
  \exp
  \left\lbrack
    - \frac{2}{3} e^{2\alpha + 2\beta_+}
    \left(
      e^{2\sqrt{3}\beta_-} + e^{-2\sqrt{3}\beta_-}
    \right)
  \right\rbrack,
  \quad j_- = 0.
\end{equation}
One sees that the equation of continuity holds and there is
a conserved current. The plane wave part of the wave
function (\ref{eq:solution-7}) propagates to the direction
of $\beta_+$ axis only. At the limit of $\beta_+ \rightarrow
- \infty$, the value of the probability density approaches a
certain constant, while converges to zero as $\beta_+
\rightarrow \infty$ or $\beta_- \rightarrow \pm\infty$.

\subsection{The Bianchi type-VIII model}

The metric of the Bianchi type-VIII model is expressed as
Eq.  (\ref{eq:metric}) and the 1-forms are given
by \cite{rf:MacCallum}
\begin{equation}
  \left\{
    \begin{array}{l}
      \chi^1 = \cosh x^2 \cos x^3 dx^1 - \sin x^3 dx^2, \\
      \chi^2 = \cosh x^2 \sin x^3 dx^1 + \cos x^3 dx^2, \\
      \chi^3 = \sinh x^2 dx^1 + dx^3.
    \end{array}
  \right.
  \label{eq:VIII-basis}
\end{equation}
The structure constants are
\begin{equation}
  C^1_{23} = - C^1_{32} = -1 ,\quad
  C^2_{31} = - C^2_{13} = -1, \quad
  C^3_{12} = - C^3_{21} = 1
  \label{eq:str-c-8}
\end{equation}
and the rest are zero.  The anisotropic potential becomes
\begin{equation}
  V_{\rm VIII} - 1 = \frac{1}{3} \left(
    e^{-8\beta_+}
    - 2 e^{4\beta_+}
    + 4 e^{-2\beta_+} \cosh(2\sqrt{3}\beta_-)
    + 2 e^{4\beta_+} \cosh(4\sqrt{3}\beta_-)
  \right) \label{eq:anisotropic-p8}.
\end{equation}
However, since the anisotropic potential
(\ref{eq:anisotropic-p8}) is a very complicated expression,
we cannot present the Dirac type equation of the form as Eq.
(\ref{eq:Dirac}).

\section{The behavior near the singularity}

It is interesting for us to discuss the behavior of the wave function 
near the singularity. Misner \cite{rf:Misner}discussed the Bianchi type-
IX (Mixmaster) model which is represented by the classical dynamical 
system where an imaginary particle moves in the anisotropic potential of 
the approximated regular triangle and regarded the dynamical system as 
the billiard in the $\beta_+$-$\beta_-$ plane.  In this way Misner's 
picture is very helpful to discuss the universe through the anisotropic 
parameters.

The vacuum Bianchi type-I universe is so-called Kasner Universe.  The
evolution of the model in classical region presents an anisotropically
expanding universe in which two axises are expanding and the other is
contracting. While, in quantum region the quadratic Hamiltonian is
derived by the ADM approach and the evolution of the quantized
universe is presented by the Wheeler-DeWitt equation like Klein-Gordon
type.  There is the problem that the probability density can become
negative by the Klein-Gordon approach.

Let us consider to change the point of view to our formalism.  Near
the singularity ($\alpha\rightarrow - \infty$), the Dirac type
equations for the Bianchi Class A models are reduced to that of
Bianchi type-I
\begin{equation}
  i \frac{\partial\Psi}{\partial\alpha}
  = 
  \left(
    \sigma_1 \hat{p}_+
    + \sigma_2 \hat{p}_-
  \right) \Psi
  \equiv \hat{h}^\prime  \Psi.
  \label{eq:hamiltonian-singularity}
\end{equation}
This dynamical system describes a massless imaginary
particle moving in zero potential. The equations of motion
in the Heisenberg representation for $\hat\beta_+$ and
$\hat\beta_-$ are
\begin{equation}
  \frac{\partial\hat\beta_+}{\partial\alpha}
  =
  i \left\lbrack \hat{h}^\prime , \hat\beta_+ \right\rbrack 
  =
  i \sigma_1 \left\lbrack \hat{p}_+, \hat\beta_+ \right\rbrack
  = \sigma_1
\end{equation}
and
\begin{equation}
  \frac{\partial\hat\beta_-}{\partial\alpha} 
  =
  i \left\lbrack \hat{h}^\prime , \hat\beta_- \right\rbrack
  =
  i \sigma_2 \left\lbrack \hat{p}_-, \hat\beta_- \right\rbrack 
  = \sigma_2,
\end{equation}
where the commutative relation $\lbrack \hat\beta_b ,
\hat{p}_a \rbrack = i\delta_{ab}\; (a,b=+,-)$ is used.  It
is similar to the {\it Zitterbewegung} \cite{rf:Dirac1} (a
rapidly oscillating motion) of electron, so that the
imaginary particle moves a trembling motions near the
singularity of the universe.  This behavior in quantum
mechanics with factorized Hamiltonian is very different from
that for the Wheeler-DeWitt equation. We emphasize here that
this kind of behavior cannot be obtained from the
Wheeler-DeWitt equation.

Here, we set an orbital angular momentum $\hat{m}^\prime$ as
\begin{equation}
  \label{eq:OAM}
  \hat{m}^\prime = \hat\beta_+ \hat{p}_- - \hat\beta_- \hat{p}_+.
\end{equation}
The Heisenberg equation of motion for the orbital angular
momentum is
\begin{equation}
  \frac{d\hat{m}^\prime}{d\alpha} 
  =
  i \left\lbrack \hat{h}^\prime , \hat{m}^\prime \right\rbrack
  =
  \sigma_1 \hat{p}_- - \sigma_2 \hat{p}_+.
\end{equation}
We have, furthermore,
\begin{equation}
  \label{eq:SAM}
  \frac{d \sigma_3}{d\alpha}
  =
  i \left\lbrack \hat{h}^\prime , \sigma_3 \right\rbrack
  =
  2 \sigma_2 \hat{p}_+ - 2\sigma_1 \hat{p}_-.
\end{equation}
The quantity $\hat{m}^\prime + \frac{1}{2} \sigma_3$ is a constant of
motion. This result one can interpret by saying a massless imaginary
particle has a `spin-like' angular momentum $\frac{1}{2}
\sigma_3$. The Dirac type equation (\ref{eq:hamiltonian-singularity})
is a similar type to the Weyl equation, which has two-dimension of the
spatial part here. We know that the Weyl equation describes the
behavior of the massless neutrino with a half spin. On the analogy of
this characteristic with the Weyl equation, the massless imaginary
particle would have a half spin.  Accordingly, we think that the Dirac type
equation for the Bianchi type-I model describes the universe as a
massless imaginary particle with `spin-like' degree of freedom. Thus,
the `spin-like' degrees of freedom cause the universe to go through an
early quantum stage of evolution with agitated
anisotropy-oscillation. However, the physical interpretation of the
`spin-like' degree of freedom is not comprehensible yet.   In
conclusion, we obtain that the Dirac type equation for the Bianchi
type-I model shows the interesting results on quantum behavior, unlike
Misner's negative conclusion that the status of the wave function
remains classical near the initial singularity \cite{rf:Misner}.

\section{Conclusions and Discussion}

We have investigated the first-order equation using the
Pauli matrices by factorizing the Wheeler-DeWitt equation
for the vacuum Bianchi Class A models except for the Bianchi
type-IX. We have derived the solution expressed in terms of
two-component spinor on the Bianchi type-I, II, VI$_0$ and
VII$_0$ models. It has been shown that the probability density
becomes positive-definite and the equation of continuity
holds with a conserved current.  However, we could not find
the solution to the Bianchi type-VIII model because the form
of anisotropic potential is very complicated expression.  It
is to be noted that we could not normalize the wave function
and the probability density for each model. In the Bianchi
type-II model, it is able to factorize the quadratic
Hamiltonian of the Wheeler-DeWitt equation into self-adjoint
Hamiltonians other than the Hamiltonian
(\ref{eq:H-II}). However, we cannot extract physically
interesting solutions to the Dirac type equations by such
Hamiltonians. The similar results are obtained for the
Bianchi type-VI$_0$ and type-VII$_0$ models.

The Dirac type equations for the Bianchi type-II, VI$_0$ and
VII$_0$ model are reduced to that for the Bianchi type-I
model near the singularity ($\alpha\rightarrow -
\infty$). The quantized universe near the singularity is
expressed by the plane wave solution with two-component
spinor and the anisotropic parameters $\hat\beta_+$ and
$\hat\beta_-$ make the motion of Zitterbewegung. The
Zitterbewegung can be related to the `spin-like' degree of
freedom and thus the `spin-like' degree of freedom causes
the universe to go necessarily through an early quantum
stage of evolution with agitated anisotropy-oscillation. The
`spin-like' degree of freedom which does not appear when we
deal with the Wheeler-DeWitt equation for the Bianchi models
may play an important role when we discuss the anisotropy of
the universe in the initial stage. In such a sense, the
Zitterbewegung is a significant subject how the big-bang or
the inflation is related. To solve the problem of the
initial singularity we must find out the physical
interpretation of the `spin-like' degree of freedom newly
found by the Dirac factorization.

As mentioned in \S 1, our method cannot be applied to the
Bianchi type-IX model. Because, the anisotropic potential of
the Bianchi type-IX $V_{\rm IX}$ is expressed in terms of
\begin{equation}
  V_{\rm IX} 
  = \frac{1}{3} e^{-8\beta_+}
  - \frac{4}{3} e^{-2\beta_+}\cosh 2\sqrt{3}\beta_-
  + \frac{2}{3} e^{4\beta_+}(\cosh 4\sqrt{3}\beta_- - 1)
  + 1 >0 .
  \label{eq:potential-V9}
\end{equation}
The quantity $v^2 = V_{\rm IX} - 1$ takes negative value as
well as positive value here. It means that the quantity $v =
\sqrt{V_{\rm IX} - 1}$ which appears in the factorized
theory turns out to be imaginary as well as real. Therefore,
we cannot factorize the quadratic Hamiltonian into a fixed
form. Hence, we must not use the Pauli matrices but the
Dirac matrices. To remedy this defect, we notice the fact
that the anisotropic potential $V_{\rm IX}$ itself is
non-negative, and carry out the Dirac procedure of
factorization with anticommuting $\gamma^\mu$ ($\mu =
0,\,1,\,2,\,3$) and $\gamma_5$ matrices.  One possibility is
to make the factorization in the form
\begin{equation}
  \left(
    \gamma^0 p_\alpha 
    + \gamma^1p_+ 
    + \gamma^2p_-
    + \gamma^3 v^\prime e^{2\alpha}  
    + \gamma_5 e^{2\alpha} 
  \right)\Psi = 0,
  \label{eq:factorized-H9}
\end{equation}
where we set $v^\prime = \sqrt{V_{\rm IX}}$.  However, as we
have seen, because the anisotropic potential $V_{\rm IX}$ is
a very complicated expression, we could not have found out the
solution to the Dirac type equation (\ref{eq:factorized-H9})
with the factorized Hamiltonian.

We compare our formalism with the Dirac-Square-Root (DSR)
formalism \cite{rf:Mallett,rf:Kim-Oh}.  On the DSR, it is
postulated that the factorized Dirac type equation for the
Bianchi type-I model is expressed as
\begin{equation}
  \label{eq:dsr}
  \left(
    i \frac{\partial}{\partial\alpha}
    - i\sigma_1 \frac{\partial}{\partial\beta_+}
    - i\sigma_2 \frac{\partial}{\partial\beta_-}
  \right) \Psi(\alpha,\beta_+,\beta_-) = 0.
\end{equation}
The Dirac type equation (\ref{eq:dsr}) then implies that the
Wheeler-DeWitt equation for the Bianchi type-I model yields
\begin{equation}
  \label{WW-1}
  \left(
    - \frac{\partial^2}{\partial\alpha^2}
    + \frac{\partial^2}{\partial\beta_+^2}
    + \frac{\partial^2}{\partial\beta_-^2}
  \right) \Psi(\alpha,\beta_+,\beta_-) = 0,
\end{equation}
for each component of the wave function. According to
Mallett \cite{rf:Mallett}, 
it is postulated that the Dirac type equation is expressed
as
\begin{equation}
    \left(
    i \frac{\partial}{\partial\alpha}
    - i\sigma_1 \frac{\partial}{\partial\beta_+}
    - i\sigma_2 \frac{\partial}{\partial\beta_-}
    + iW(\alpha) 
  \right) \Psi(\alpha,\beta_+,\beta_-) = 0,
  \label{eq:M-K-O}
\end{equation}
where $W(\alpha)$ is an unknown function depended on $\alpha$
only. In Kim and Oh \cite{rf:Kim-Oh} paper, it has been
applied the first-order equation (\ref{eq:M-K-O}) to the
Bianchi type-IX model.

However, it is not valid to apply the DSR formalism to the
Wheeler-DeWitt equation including the anisotropic potential
term like (\ref{eq:potential-V9}) and to factorize in
the form of the Dirac type equation (\ref{eq:M-K-O}).
Because, it is inconsistent with the original Wheeler-DeWitt
equation in the point that the differential operators with
the anisotropic parameters $\beta_+$ and $\beta_-$ do not
act on the unknown function $W (\alpha)$.  The incorrect
equation yields if one assume the equation (\ref{eq:M-K-O})
when the Wheeler-DeWitt equation has the anisotropic
potential term included the anisotropic parameters.  While,
our formalism is valid to apply to the Bianchi Class A
models because it is able to factorize the Wheeler-DeWitt
equation included the anisotropic potential term into the
Dirac type equation (\ref{eq:Dirac}).  This is the extended
DSR formalism.

When we yield the original expression from the Dirac type
equation discussed by our formalism and the iterating
equation, we also obtain the extra terms which include the
time parameter and the anisotropic parameters. Regarding the
extra terms as a new constraint to the wave function and
solving the constraint, we cannot get solutions which
satisfy the Dirac type equation. As a consequence, we only
discuss the Dirac type equation without imposing the
constraint condition on the wave function.

\section*{Acknowledgments}
The author is grateful to Profs. Tetsuya Hara and Ikuo S. Sogami for the
motivation of this work and helpful discussion.

\def\PTP#1{Prog.\ Theor.\ Phys.\ \andvol{#1}}
\def\JPSJ#1{J.~Phys.\ Soc.\ Jpn.\ \andvol{#1}}
\def\PR#1{Phys.\ Rev.\ \andvol{#1}}
\def\PRL#1{Phys.\ Rev.\ Lett.\ \andvol{#1}}
\def\PL#1{Phys.\ Lett.\ \andvol{#1}}
\def\NP#1{Nucl.\ Phys.\ \andvol{#1}}
\def\JMP#1{J.~Math.\ Phys.\ \andvol{#1}}
\def\IJMP#1{Int.\ J.~Mod.\ Phys.\ \andvol{#1}}
\def\CMP#1{Commun.\ Math.\ Phys.\ \andvol{#1}}
\def\JP#1{J.~of Phys.\ \andvol{#1}}
\def\ANN#1{Ann.\ of Phys.\ \andvol{#1}}
\def\NC#1{Nouvo Cim.\ \andvol{#1}}

\end{document}